\documentclass[sigconf]{acmart}

\usepackage{xspace} 
\usepackage[inline]{enumitem}
\usepackage{amsmath}
\usepackage{multirow}
\usepackage{multicol}

\def\BibTeX{{\rm B\kern-.05em{\sc i\kern-.025em b}\kern-.08emT\kern-.1667em\lower.7ex\hbox{E}\kern-.125emX}}

\newlist{inlinelist}{enumerate*}{1}
\setlist*[inlinelist,1]{%
  label=(\roman*),
}
\copyrightyear{2019} 
\acmYear{2019} 
\setcopyright{acmcopyright}
\acmConference[ICTIR '19]{The 2019 ACM SIGIR International Conference on the Theory of Information Retrieval}{October 2--5, 2019}{Santa Clara, CA, USA}
\acmBooktitle{The 2019 ACM SIGIR International Conference on the Theory of Information Retrieval (ICTIR '19), October 2--5, 2019, Santa Clara, CA, USA}
\acmPrice{15.00}
\acmDOI{10.1145/3341981.3344240}
\acmISBN{978-1-4503-6881-0/19/10}

\newcommand{\modelname}{CATAPE\xspace}

\newcommand{\partitle}[1]{\vspace{2mm}\noindent\textbf{#1}}

\begin{document}
\title{Category-Aware Location Embedding for \\ Point-of-Interest Recommendation}

\author{Hossein A. Rahmani}
\affiliation{%
  \institution{University of Zanjan}
}
\email{srahmani@znu.ac.ir}

\author{Mohammad Aliannejadi}
\affiliation{%
  \institution{Universit\`a della Svizzera italiana (USI)}
}
\email{mohammad.alian.nejadi@usi.ch}

\author{Rasoul Mirzaei Zadeh}
\affiliation{%
  \institution{University of Zanjan}
 }
\email{mirzaei.rasoul@znu.ac.ir}

\author{Mitra Baratchi}
\affiliation{%
  \institution{Leiden University}
 }
\email{m.baratchi@liacs.leidenuniv.nl}

\author{Mohsen Afsharchi}
\affiliation{%
  \institution{University of Zanjan}
 }
\email{afsharchi@znu.ac.ir}

\author{Fabio Crestani}
\affiliation{%
  \institution{Universit\`a della Svizzera italiana (USI)}
}
\email{fabio.crestani@usi.ch}

%
\renewcommand{\shortauthors}{Hossein A. et al.}
%
\begin{abstract}
Recently, Point of interest (POI) recommendation has gained ever-increasing importance in various Location-Based Social Networks (LBSNs). With the recent advances of neural models, much work has sought to leverage neural networks to learn neural embeddings in a pre-training phase that achieve an improved representation of POIs and consequently a better recommendation. However, previous studies fail to capture crucial information about POIs such as categorical information.

In this paper, we propose a novel neural model that generates a POI embedding incorporating sequential and categorical information from POIs. Our model consists of a check-in module and a category module. The check-in module captures the geographical influence of POIs derived from the sequence of users' check-ins, while the category module captures the characteristics of POIs derived from the category information. To validate the efficacy of the model, we experimented with two large-scale LBSN datasets. Our experimental results demonstrate that our approach significantly outperforms state-of-the-art POI recommendation methods.
\end{abstract}


\maketitle

\section{Introduction}
With the growing popularity of Location-Based Social Networks (LBSNs) such as Foursquare and Yelp users share their check-in records and the experiences they had while visiting different points of interest (POIs), such as restaurants, shopping malls and museums. The availability of such a myriad of data has opened various research opportunities in POI recommendation~\cite{liu2016exploring,DBLP:journals/tois/AliannejadiC18,liu2017experimental}. For instance, much work has been done with a focus on modeling sequential check-in information. Since sequential check-ins reveal a wealth of latent information about POIs and user preferences before or after visiting specific POIs, relevant studies argue that modeling the sequential check-in information is crucial for POI recommendation~\cite{liu2016exploring,feng2017poi2vec}.

Recent work has shown the effectiveness of pre-trained POI embeddings in improving the recommendation performance~\cite{erhan2010does,he2017neural}. In particular, the main idea behind this approach is to learn a representation of POIs based on a large amount of check-in data and use the pre-trained embeddings as initial values of the latent representations of POIs in conventional recommendation models~\cite{chang2018content,zhao2017geo}.
Another line of research points out the importance of POI categories as they convey useful information regarding users' interests and habits~\cite{DBLP:journals/tois/AliannejadiC18,DBLP:conf/sigir/ZhangC15, DBLP:conf/ictir/AliannejadiRC18}. Therefore, it is critical to incorporate this information while learning the POI embeddings. However, no effort has been done yet in this direction.


In this paper, we aim to study the effect of categorical information on the performance of our proposed POI embedding model. To this end, we propose a two-phase embedding model that captures the sequential check-in patterns of users together with the categorical information existing on LBSN data. Our model, called \textbf{Cat}egory-\textbf{A}ware \textbf{P}OI \textbf{E}mbedding (\modelname), learns a high dimensional representation of POIs based on two data modalities, i.e., check-in sequence and POI categories. More specifically, the contributions of this paper can be summarized as follows:
\begin{inlinelist}
    \item We shows that the characteristics of POIs is important in POI embeddings;
    \item We propose a novel category-aware POI embedding model that utilizes a user's check-in sequence information, as well as, POI categorical information;
    \item We evaluate our proposed method of \modelname on two large-scale datasets, comparing the performance of our model with state-of-the-art approaches.
\end{inlinelist}
The experimental results demonstrate the effectiveness of our POI embedding, outperforming state-of-the-art POI recommendation models significantly. Furthermore, we show that incorporating the categorical information into the embedding enables \modelname to capture users' interests and POIs' characteristics more accurately.

\section{Related Work}
In this section, we give a review of previous POI recommendation models. Modeling contextual information such as geographical, temporal, categorical, and social in POI recommendation systems has been proven to be a necessary step to improve the quality of recommendation results \cite{cheng2012fused,yuan2013time}. To develop context-aware applications, \citet{li2015rank} modeled the task of recommending POIs as the problem of pair-wise ranking, and used the geographical information to propose a ranking-based geographical factorization method.
\citet{ye2011exploiting} argued that users' check-in behavior is affected by the spatial influence of locations and proposed a unified location recommender system incorporating spatial and social influence to address the data sparsity problem. However, this method does not consider the spatial information based on each individual user, but rather models it based on all users' check-in distribution. \citet{cheng2012fused} proposed a multi-center Gaussian model to capture users' movement pattern as they assumed users' movements happen around several centers. 

With the increasing interest in modeling sequential patterns in other fields such as NLP and the successful use of these approaches in POI recommendation to consider the relation between the sequence of visited locations, much attention has recently been drawn towards modeling sequential patterns in POI recommendation~\cite{liu2016exploring,chang2018content,feng2017poi2vec,zhao2017geo}. \citet{liu2016exploring} adopted the word2vec framework to model the check-in sequences capturing the sequential check-in patterns. \citet{zhao2017geo} proposed a model for POI recommendation with attention to the fact that check-in sequences depend on the day of the week, for instance, work on weekday and entertainment on weekend. Moreover, \citet{feng2017poi2vec}, presented a latent representation model that is able to incorporate the geographical influence. However, they do not incorporate the contextual information in their model and do not consider the characteristics of POIs. More recently, \citet{chang2018content} proposed a content-aware POI embedding model which incorporated the textual content of POIs into the embedding model. Our work, in contrast, focuses on incorporating categorical information into the POI embedding model to model the characteristics of POIs.

\section{Proposed Method}
In this section, we propose a category-aware POI embedding model called \modelname, which captures both the geographical and categorical information of POIs. Our model generates a high dimensional representation of POIs, which is then plugged into a POI recommendation model to produce a recommendation list. In the following, we first describe an overview of our embedding model and further explain how it is implemented and how the generated POI embeddings are incorporated into a recommender system.

Formally, let $L=\{l_1, l_2, ..., l_n\}$ be the set of POIs and $C=\{c_1, c_2, ..., c_d\}$ be the set of categories where $c_i$ determines the category label that can be associated with any $l_j$. Moreover, let $S_{l_j}=(s_1, s_2, ..., s_q)$ be the sequence of check-ins that occurred before and after $l_j$ where $s_i$ determines the check-in to POI $l_j$. 

\modelname estimates the probability $p(V = 1 | l, c, S_l; L, C)$, where $V$ is a binary random variable indicating whether the POI $l$ with category $c$ should be predicted as part of the $S_l$ POI sequence. As mentioned earlier, $S_l$ provides the sequential context of a check-in. The POI recommendation probability is estimated as follows:
\begin{equation}
p(V = 1 | l, c, S_l; L, C) = \psi_{emb}(\phi_{ch}(l, S_l), \phi_{ct}(l, c))~,
\end{equation}
where $\phi_{ch}$ and $\phi_{ct}$ denote Check-in module and Category module, respectively. $\psi$, on the other hand, is a POI recommender component that takes the POI embeddings learned by $\phi_{ch}$ and $\phi_{ct}$ and generates a POI classification probability. 



\partitle{Check-in module.}
The visited POIs by users, which are consecutive in the user's check-in context, are geographically influenced by each other. The check-in module proposed in the study by \cite{liu2016exploring} employs the Skip-gram based POI embedding model to capture the user check-in sequence to POIs. For every target POI $l\in{L}$, $W(l, S_l)$ generates the context of $l$, extracted from $S_l$. It includes the POIs visited before and after $l$ based on the pre-defined window size. The check-in module trains the POI embedding vector by maximizing the objective function of Skip-gram model \cite{mikolov2013efficient} in the following way:
\begin{equation}
    \phi_{ch}(l, S_l) = arg \ max \prod_{l\in{L}}[\prod_{l_w\in{W(l,  S_l)}}p(l_w|l)]~.
    \label{eq:OF-skipgrapm}
\end{equation}
For a given POI $l$, the probability $p(l_w|l)$ of context POI $l_w$ is computed using the soft-max function as follows:
\begin{equation}
   p(l_w|l) = \frac{\exp{(v^T_w.v_l})}{\sum_{k=1}^{|L|}\exp{(v^T_k.v_l})}~,
\end{equation}
where $v_l\in{\mathbb{R}^{D\times1}}$ and $v_w\in{\mathbb{R}^{D\times1}}$ are the latent vectors of the target and context POI, respectively. $D$ is the dimensionality of the latent space and $|L|$ is the number of POIs in the set $L$.

As the size of $L$ in Eq. \eqref{eq:OF-skipgrapm} is typically very large, to improve and speed up the process of optimization we adopt a negative sampling technique \cite{mikolov2013efficient}. Thus, the loss function (i.e., negative log) for check-in module is defined as follows:
\begin{equation}
\begin{split}
    loss_{checkin}=-\sum_{l\in{L}}\sum_{l_w\in{W(l)}}(\log_{}{\sigma(v_w^T.v_l)} \\
    +\sum_{e\in{L_{neg}}}\log_{}{\sigma(-v_e^T.v_l)})~,
    \label{eq:loss_checkin}
\end{split}
\end{equation}
where $\sigma(\cdot)$ is the sigmoid function and $L_{neg}$ is the set of negative POI samples, i.e., the POIs that do not appear in the context window of $l$.

\partitle{Category module.}
We propose $\psi_{ct}$ inspired  by the idea of word2vec~\cite{mikolov2013efficient} where we consider the categories of POIs visited by a user as a ``sentence'' and every single category as a ``word.'' Therefore, $\psi_{ct}$ is estimated as follows:
\begin{equation}
    \phi_{ct}(l, c) = arg \ max \prod_{l_i\in{L}}\prod_{c_j\in{C_i}}\prod_{c_q\in{Q(c_j)}}p(c_q|c_j, l_i)~,
\end{equation}
where $C_i$ is the category information of POI $l_i$, $c_j$ is the target category, and $Q(c_j)$ is the categories visited before and after category $c_j$ based on  a pre-defined window size. The probability function $p(c_q|c_j, l_i)$ is estimated using the soft-max function as follows:
\begin{equation}
    p(c_q|c_j, l_i) = \frac{\exp{(v_q^T\cdot{\hat{v_j}})}}{\sum_{n=1}^{|C|}\exp{(v_n^T\cdot{\hat{v_j}})}}~,
\end{equation}
where $\hat{v_j} = l_i \oplus c_j$ and $\oplus$ is the concatenation function and $|C|$ is the number of categories in the set $C$. Similar to \eqref{eq:loss_checkin}, for efficiency, we formulate the loss function of the category module $loss_{category}$ using the negative sampling technique as follows:
\begin{equation}
\begin{split}
    loss_{category}=-\sum_{l_i\in{L}}\sum_{c_j\in{C_i}}\sum_{c_q\in{Q(c_j)}}(\log_{}{\sigma(v_q^T.\hat{v_j})} \\
    +\sum_{r\in{G_{neg}}}\log_{}{\sigma(-v_r^T.\hat{v_j})})~,
\end{split}
\end{equation}
where $\sigma(\cdot)$ is the sigmoid function and $G_{neg}$ is the set of negatively sampled categories.
Finally, we combine the Check-in and Category modules. The final objective function of our category-aware POI embedding model \modelname is:
\begin{equation}
    \psi_{emb} = \phi_{ch}(l, S_l) + \phi_{ct}(l, c)~.
\end{equation}
\modelname maximizes the final objective function when it simultaneously learns the geographical and categorical influence of POIs.

\begin{figure}
    \vspace{-0.0cm}
    \centering
    \includegraphics[width=0.7\columnwidth]{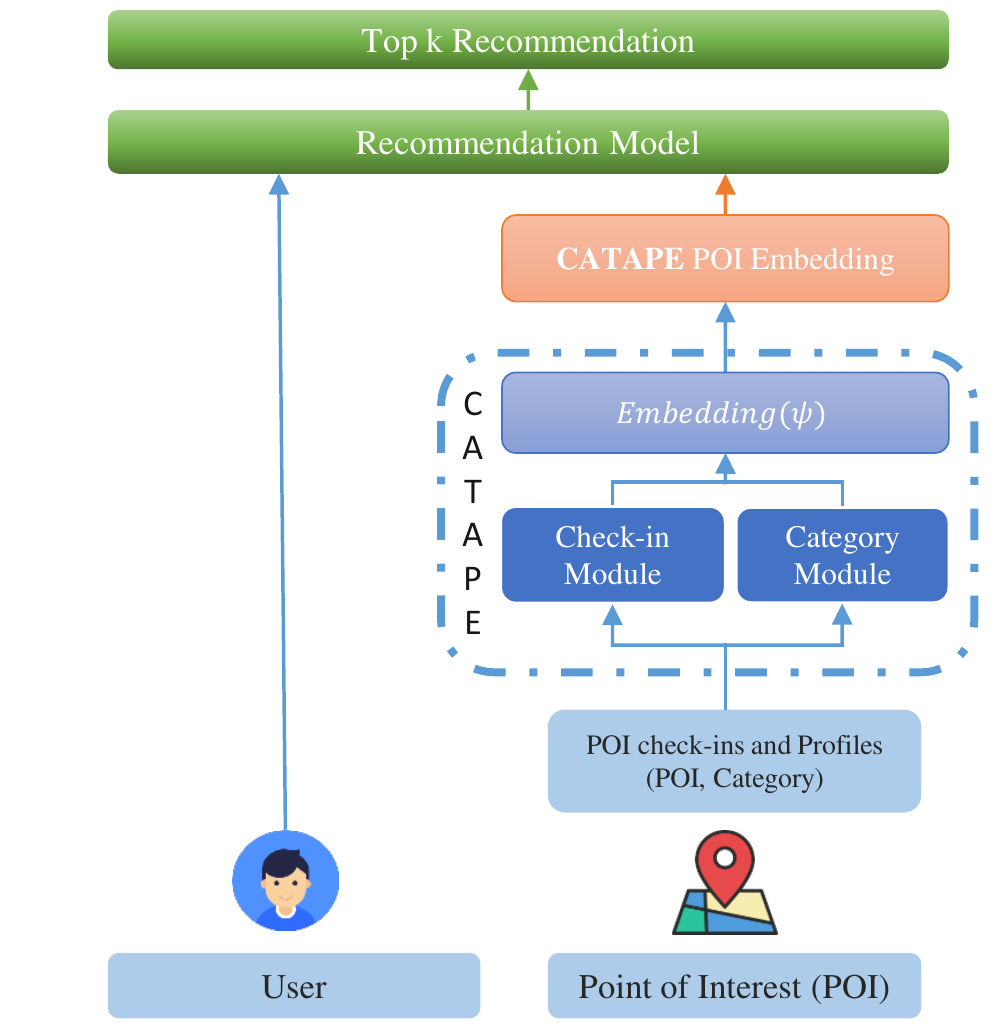}
    \caption{The workflow of utilizing pre-trained POI embedding from \modelname to produce top k recommendation list.}
    \label{fig:model}
\end{figure}

\partitle{System overview.}
As we discussed earlier, the POI embeddings are learned as part of a classification model where the model determines a missing POI in a sequence of visited POIs (i.e., Skip-gram model). This is not an actual POI recommendation setting which considers users and POI at the same time. In the next step, we extract the learned POI embeddings from the trained network to feed to a POI recommendation model. Figure~\ref{fig:model} illustrates our proposed workflow of recommendation using pre-trained POI embeddings. As seen, based on the check-in data in the training set, \modelname learns the high-dimensional POI embeddings. Then, the learned POI embedding is fed to a recommender model which is able to utilize this information to provide accurate recommendation. In this work, we used Metric Factorization \cite{2018arXiv180204606Z} as the recommender model since it is a state-of-the-art recommender model that is able to use pre-trained POI embeddings. This model takes our pre-trained POI embeddings and learns user embeddings to recommend top $k$ recommendation list. 

\section{Experiments}
In this section, we evaluate the performance of \modelname in comparison with a set of state-of-the-art POI recommendation models.

\subsection{Experimental Setup}
\partitle{Dataset.}
We evaluate the performance of \modelname on two real-world data, Yelp \cite{liu2017experimental} (access date: Feb 2016) and Gowalla \cite{yuan2013time} were collected between Feb. 2009 and Oct. 2010. The Yelp dataset consists of $357,410$ check-ins made by $6,254$ users on $21,240$ POIs and $507$ unique categories. The Gowalla dataset contains $456,988$ check-ins made by $10,162$ users at $24,250$ POIs with $417$ categories. For every user, we consider the first $80\%$ of check-ins in chronological order as the training set, followed by the last $20\%$ check-ins as test.

\partitle{Evaluation metrics.}
We measure the effectiveness of the recommendation task using two standard recommendation evaluation metrics: Precision@$k$ for the top $k$ recommended POIs and Recall@$k$ for the top $k$ recommended POIs. In Table \ref{tab:pre_yelp}, statistically significant results are shown, which achieved by performing a two-tailed paired t-test at a $95\%$ confidence interval ($p<0.05$).

\partitle{Compared methods.}
The main focus of our work is capturing POI check-in sequence and category context\footnote{In our experiments, we set the embedding dimension and pre-defined window size to 100 and 4, respectively.}. These context information can be derived from a pure check-in dataset without any additional data modality (e.g, user comments, user profiles). Respectively, to be fair we have chosen a number of baselines that also perform on check-in datasets. These models are, \cite{ye2011exploiting}, \cite{cheng2012fused}, \cite{li2015rank}, \cite{zhao2017exploiting} that can capture geographical, social, and categorical context. It should be mentioned that, among the previous works which involve learning embeddings during a pre-training phase, the one proposed in \cite{chang2018content} considers availability of textual context provided in user reviews to incorporate characteristics of POIs. Therefore, comparison with this model is out of our scope. Our model can be considered complementary to these models. We compare the performance of \modelname with the following models:
\begin{itemize}[leftmargin=*]
    \item \textbf{USG} \cite{ye2011exploiting} takes advantage of three modules of user-based CF, social influence, and geographical information.
    \item \textbf{MGMPFM} \cite{cheng2012fused} combines geographical influence with Probabilistic Factorization Model (PFM), assuming a Multi-Center Gaussian Model (MGM) of the probability of a user's check-in behavior.
    \item \textbf{BPRMF} \cite{rendle2009bpr} adopts a Bayesian criterion to directly optimize for personalized rankings based on users' implicit feedback.
    \item \textbf{RankGeoFM} \cite{li2015rank} is a state-of-the-art ranking-based geographical factorization method. It incorporates the geographical information in a latent ranking model.
    \item \textbf{HGMF} \cite{zhao2017exploiting} is a state-of-the-art hierarchical geographical matrix factorization model to utilize the hierarchical structures of both users and POIs with categorical information for POI recommendation.
    \item \textbf{Metric Factorization} \cite{2018arXiv180204606Z} places users and POIs in a low dimensional space and measures their explicit similarity using Euclidean distance.
    \item \textbf{CATAPE-NoCat} is a variation of \modelname in which we remove the category module from the model. Therefore, it is trained using only the check-in module.
\end{itemize}

\begingroup
\setlength{\tabcolsep}{2pt}
\begin{table*}[t]
    \vspace{-0.0cm}
  \caption{Performance comparison with baselines in terms of Precision@$k$ and Recall@$k$ for $k \in \{5,10,20\}$ on Gowalla and Yelp. The superscripts $\dagger$ and $\ddagger$ denote significant improvements compared to baselines and CATAPE-NoCat, respectively ($p < 0.05$).}
  \label{tab:pre_yelp}
  \vspace{-0.3cm}
  \begin{tabular}{lllllllcllllll}
    \toprule
    \multirow{2}{*}{\textbf{Method}} & \multicolumn{6}{c}{\textbf{Yelp}} && \multicolumn{6}{c}{\textbf{Gowalla}} \\
    \cmidrule{2-7} \cmidrule{9-14}
    & P@5 & P@10 & P@20 & R@5 & R@10 & R@20 && P@5 & P@10 & P@20 & R@5 & R@10 & R@20\\
    \midrule
    USG & 0.0282 & 0.0244 & 0.0197 & 0.0281 & 0.0523 & 0.0753 && 0.0502 & 0.0471 & 0.0413 & 0.0517 & 0.0568 & 0.0625 \\
    MGMPFM & 0.0197 & 0.0173 & 0.0136 & 0.0211 & 0.0293 & 0.0493 && 0.0281 & 0.0215 & 0.0197 & 0.0263 & 0.0291 & 0.0319 \\
    BPRFM & 0.0285 & 0.0221 & 0.0185 & 0.0296 & 0.0361 & 0.0599 &&  0.0493 & 0.0443 & 0.0342 & 0.0497 & 0.0529 & 0.0581 \\
    RankGeoFM & 0.0421 & 0.0362 & 0.0292 & 0.0392 & 0.0673 & 0.0838 && 0.0567 & 0.0501 & 0.0492 & 0.0591 & 0.0642 & 0.0718 \\
    HGMF & 0.0532 & 0.0491 & 0.0401 & 0.0478 & 0.0702 & 0.0915 &&  0.0798 & 0.0711 & 0.0683 & 0.0715 & 0.0773 & 0.0819 \\
    Metric Factorization & 0.0593 & 0.0552 & 0.0481 & 0.0533 & 0.0782 & 0.0974 &&  0.0821 & 0.0782 & 0.0717 & 0.0784 & 0.0814 & 0.0862 \\
    \midrule
    CATAPE-NoCat & 0.0641$^\dagger$ & 0.0613$^\dagger$ & 0.0568$^\dagger$ & 0.0589 & 0.0831$^\dagger$ & 0.1013$^\dagger$ &&  0.0892$^\dagger$ & 0.0828$^\dagger$ & 0.0784$^\dagger$ & 0.0815$^\dagger$ & 0.0898$^\dagger$ & 0.0979$^\dagger$ \\
    \modelname & \textbf{0.0702}$^{\dagger\ddagger}$ & \textbf{0.0692}$^{\dagger\ddagger}$ & \textbf{0.0631}$^{\dagger\ddagger}$ & \textbf{0.0621}$^{\dagger\ddagger}$ & \textbf{0.0881}$^{\dagger}$ & \textbf{0.1121}$^{\dagger\ddagger}$ &&  \textbf{0.0924}$^{\dagger\ddagger}$ & \textbf{0.0894}$^{\dagger\ddagger}$ & \textbf{0.0813}$^{\dagger\ddagger}$ & \textbf{0.0872}$^{\dagger}$ & \textbf{0.0953}$^{\dagger\ddagger}$ & \textbf{0.1283}$^{\dagger\ddagger}$ \\
  \bottomrule
\end{tabular}
\vspace{-0.3cm}
\end{table*}
\endgroup

\vspace{-0.3cm}
\subsection{Results and Discussion}
In the following section, we report the performance of \modelname, analyzing its effectiveness compared with other methods.

\partitle{Performance comparison.}
Table~\ref{tab:pre_yelp} lists the performance of \modelname, as well as, the compared methods in terms of Precision@$k$ and Recall@$k$, respectively. As seen, between the baseline methods MGMPFM has the least performance in terms of all metrics. USG outperforms the MGMPFM by $78.5\%$ in terms of Rec@10 on Yelp. The results show that the HGMF consistently achieves the best performance against USG, MGMPFM, BPRMF, and RankGeoFM considering the hierarchical structure and incorporating the categorical information as one the most effective contextual signals in the hierarchical model. However, HGMF considers the dot product of users and POIs in measuring their similarity. Among the baselines, it is seen that Metric Factorization outperforms HGMF and other methods, suggesting that using Euclidean distance is a more precise measure of similarity as opposed to dot product.

As seen in Table~\ref{tab:pre_yelp}, \modelname significantly outperforms all of the baseline methods in terms of all evaluation metrics. This indicates that the check-in module is able to learn POI latent representation by modeling the context of users' visited POIs and the sequence of POIs. Furthermore, the results suggest that incorporating category information enables \modelname to model the characteristics of the POIs more effectively.
It is worth noting that our proposed POI embedding model can be pre-trained on a large dataset of check-ins to be used in various POI recommendation models.
Also, it is seen that CATAPE-NoCat is able to outperform all the baselines significantly, indicating that learning POI embeddings only based on check-in information is able to capture complex sequential relations between POIs.

\partitle{Impact of category module.}
To show the effect of category information on the performance, we compare the performance of \modelname with \modelname-NoCat.
The results in Table~\ref{tab:pre_yelp} show that the performance of model significantly drops when we remove category information, indicating that the category information enables the model to capture the similarities between POIs more accurately. As mentioned in the literature~\cite{DBLP:journals/tois/AliannejadiC18}, category information is crucial for capturing users regular habits. For instance, a user may stop by a drive-thru coffee shop every morning, just before going to their workplace. 
Despite the performance drop, it is seen that CATAPE-NoCat is able to outperform all the baseline methods significantly in terms of all evaluation metrics. More specifically, it is seen that CATAPE-NoCat outperforms Metric Factorization by $18\%$ in terms of Pre@20 on Yelp and $13.57\%$ in terms of Rec@20 on Gowalla.
Finally, note that a similar experiment, removing the check-in module would not be possible because in the category module latent vectors of POIs, computed by the check-in module, are required.

\vspace{-0.3cm}
\section{Conclusions}
In this paper, we introduced a novel POI embedding model and demonstrated the importance of characteristics of POIs in POI embedding. Our model captures the sequential influence of POIs from check-in sequence of users, as well as, characteristics of POIs using the category information. The experimental results showed that our model contributes to improving POI recommendation performance.
\bibliographystyle{ACM-Reference-Format}
\bibliography{references}
\end{document}